\documentclass{optica-article}

\journal{opticajournal} 

\articletype{Research Article}

\usepackage{graphicx}
\usepackage{dcolumn}
\usepackage{bm}
\usepackage[utf8]{inputenc}
\usepackage[T1]{fontenc}
\usepackage{xcolor}
\usepackage{units}
\usepackage{lineno}
\usepackage{braket}
\usepackage{amsmath} 
\usepackage{bbold} 
\usepackage[normalem]{ulem}

\begin{document}
	
	
	\title{High fidelity distribution of triggered polarization-entangled telecom photons via a
		\unit[36]{km} intra-city fiber network} 
	
	
	
	\author{Tim Strobel,\authormark{1,*} Stefan Kazmaier,\authormark{1} Tobias Bauer,\authormark{2} Marlon Sch\"afer,\authormark{2} Ankita Choudhary,\authormark{3} Nand Lal Sharma,\authormark{3} Raphael Joos,\authormark{1} Cornelius Nawrath,\authormark{1} Jonas H. Weber,\authormark{1} Weijie Nie,\authormark{3} Ghata Bhayani,\authormark{3}  Lukas Wagner,\authormark{1} Andr\'e Bisquerra,\authormark{1} Marc Geitz,\authormark{4} Ralf-Peter Braun,\authormark{4} Caspar Hopfmann,\authormark{3} Simone L. Portalupi,\authormark{1} Christoph Becher,\authormark{2} and Peter Michler,\authormark{1}}
	
	\address{\authormark{1}{Institut f\"ur Halbleiteroptik und Funktionelle Grenzfl\"achen, Center for Integrated Quantum Science and Technology ($IQ^{ST}$) and SCoPE, University of Stuttgart, Allmandring 3, 70569 Stuttgart, Germany\\
		\authormark{2}Fachrichtung Physik, Universit\"at des Saarlandes, Campus E2.6, 66123 Saarbr\"ucken, Germany\\
		\authormark{3}Institute for Integrative Nanosciences, Leibniz IFW Dresden, Helmholtzstraße 20, 01069 Dresden, Germany}
		\authormark{4}Deutsche Telekom T-Labs Landgrabenweg 151, 53227 Bonn, Germany}
	
	\email{\authormark{*}t.strobel@ihfg.uni-stuttgart.de}

	\date{\today}
	
	\begin{abstract*}

		Fiber-based distribution of triggered, entangled, single-photon pairs is a key requirement for the future development of terrestrial quantum networks. In this context, semiconductor quantum dots (QDs) are promising candidates for deterministic sources of on-demand polarization-entangled photon pairs. So far, the best QD polarization-entangled-pair sources emit in the near-infrared wavelength regime, where the transmission distance in deployed fibers is limited. Here, to be compatible with existing fiber network infrastructures, bi-directional polarization-conserving quantum frequency conversion (QFC) is employed to convert the QD emission from \unit[780]{nm} to telecom wavelengths. We show the preservation of polarization entanglement after QFC (fidelity to Bell state $F_{\phi^+, conv}=0.972\pm0.003$) of the biexciton transition. As a step towards real-world applicability, high entanglement fidelities ($F_{\phi^+, loop}=0.945\pm0.005$) after the propagation of one photon of the entangled pair along a \unit[35.8]{km} field installed standard single mode fiber link are reported. Furthermore, we successfully demonstrate a second polarization-conversing QFC step back to \unit[780]{nm} preserving entanglement ($F_{\phi^+, back}=0.903\pm0.005$). This further prepares the way for interfacing quantum light to various quantum memories.
		
	\end{abstract*}

	\section{Introduction}
	
	In recent years, the research community has directed significant attention towards exploring the potential of quantum technologies. On the one hand, scientists aim at increasing the knowledge of quantum mechanics applied to specific tasks, often with the aim of reaching a distinct quantum advantage over classically-based approaches. On the other hand, more and more implementations target realizations outside of a controlled lab environment~\cite{Liu2023,Vallone2015,Ursin2007a,Sasaki2011}. One cornerstone of the second quantum revolution is the use of entanglement. Particularly intriguing for quantum communication and information science is the possibility of providing a secure communication scheme between distant nodes.\\
	Among various systems, semiconductor quantum dots (QD) emerged as an promising candidate for the generation of highly entangled photon pairs with high brightness. The widespread classical communication networks of today are largely based on optical silica fiber connections. To minimize losses in these fiber networks - an aspect which is even more relevant in quantum optical communication schemes~\cite{VanLoock2020} -, photons at telecom wavelengths (O- and C-bands) are used.\\
	For the generation of polarization-entangled photons at telecommunication wavelength, strain engineering of the well developed In(Ga)As quantum dots has been employed~\cite{Olbrich2017,Zeuner2021}, alternatively employing InAs on InP systems~\cite{Anderson2020}. Still, the highest reported degree of entanglement has been achieved with QDs emitting photons in the near-infrared (NIR)~\cite{DaSilva2021,Hopfmann2020,Huber2018c,Pennacchietti2024} far from the low absorption bands of silica fibers. To overcome this limitation quantum frequency conversion (QFC) of single NIR photons to low-loss telecom bands can be applied. QFC is a mature technology that has been shown to preserve coherence and single photon statistics~\cite{Zaske2012,Ates2012,Morrison2023}, as well as indistinguishability of single photons~\cite{Kambs2016a,Yu2015,Ikuta2013}. Moreover, polarization-preserving conversion devices have been developed enabling the conversion of polarization qubits with external device efficiencies of more than \unit[50]{\%}~\cite{Bock2018,VanLeent2022a,Arenskotter2022}. QFC in addition plays a crucial role in spectrally matching photons from independent quantum emitters by adjusting the conversion wavelength, thus, increasing their indistinguishability in two-photon interference experiments~\cite{Weber2019c,Stolk2022,Knaut2023}. Several groups have demonstrated the feasibility of out-of-the-lab experiments with single and entangled photons from various material platforms~\cite{Shi2020,Chen2021,Neumann2022,Businger2022,Arenskotter2022,Ribezzo2023,Pelet2023} including quantum dots~\cite{Yang2023,Zahidy2024}. Even with entangled photons from QDs, quantum key distribution experiments have been performed at around \unit[780]{nm} over several hundred meters~\cite{Schimpf2021,Basset2021}. Longer entanglement distribution distances have been achieved with telecom photons emitted by a QD under continuous wave pumping, showing an interesting perspective for utilizing entanglement within deployed standard telecom fibers~\cite{Xiang2019}.\\	
	In this work, we employ a bright source based on epitaxially grown droplet etching GaAs QDs embedded in a dielectric antenna design~\cite{Hopfmann2020} to generate polarization-entangled photon pairs at around $\unit[780]{nm}$. Together with record values of entanglement generation, such platform has shown recently unprecedented spin coherence time which makes these QDs also appealing for effective spin-photon interface experiments~\cite{Zaporski2023}. Here we employ pulsed resonant excitation to generate polarization entangled photon pairs from the biexciton (XX) - exciton (X) - ground (G) cascade. The XX transition is further frequency converted to telecom wavelength ($\sim\unit[1515]{nm}$), while preserving the degree of entanglement. This one photon at telecom wavelengths is sent via a $35.8\,$km field installed intra-city fiber link while keeping its entangled state. This measurement is essential for the implementation of practical quantum communication networks. Finally, it is also shown that a second QFC can be used to convert the telecom photon back to NIR~\cite{Arenskotter2022}, all while preserving the entanglement. This becomes key for the precise spectral matching to atomic transitions such as the Rb~D$_{2}$ line, or for an effective interface with a different GaAs QD spin~\cite{Zaporski2023}. This would allow for a flexible control of the photon wavelength, enabling both low loss propagation in fiber networks, and interfacing with atomic memories~\cite{Heshami2016,Neuwirth2021,Finkelstein2018,Kaczmarek2018}.
	
	\section{Results}
	
\begin{figure}
		\centering
		\includegraphics[width=0.8\linewidth]{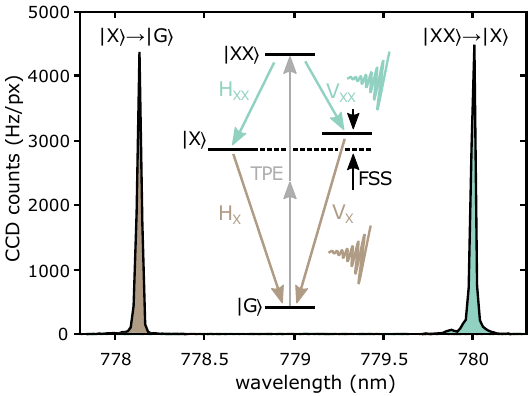}
		\caption{\textbf{Spectrum of the entangled photon source and energy diagram.} Micro-photoluminescence spectrum of the employed quantum dot, depicting the biexciton-to-exciton emission (green) and the exciton-to-ground state (brown) in two-photon excitation at $\pi$-pulse~\cite{Stufler2006}. The inset visualizes the schematic energy structure and biexciton-exciton cascade of the employed three-level system with a fine-structure splitting of $\text{FSS}=\unit[2.1]{\mu eV}$. The excitation process in two-photon excitation (TPE) is represented by gray arrows.}
		\label{fig:fig1}
	\end{figure}

	Semiconductor QDs enable the creation of polarization-entangled photon pairs in the biexciton-exciton-cascade~\cite{Akopian2006,Young2006,Hafenbrak2007}. This emission can be particularly efficient, as well as triggered, when pulsed two-photon-excitation (TPE) is employed~\cite{Jayakumar2013,Muller2014a}. Figure \ref{fig:fig1} shows the resulting fiber-coupled micro-photoluminescence spectrum in TPE and the associated inset depicts the corresponding energy diagram. Similar emission intensities of X and XX confirm the successful preparation of the biexciton state, followed by the emission of a photon pair~\cite{Muller2014a}. Thanks to the resonant TPE scheme spectral filtering using volume Bragg gratings (VBG) can be used to fully suppress  background from the TPE. Ideally the QD emits a two-photon state $\ket{\psi(t)}$ rotating between Bell-states $\phi^+=1/\sqrt{2}\left(\ket{H_{XX}H_X}+\ket{V_{XX}V_X}\right)$ and $\phi^-=1/\sqrt{2}\left(\ket{H_{XX}H_X}-\ket{V_{XX}V_X}\right)$ with a phase given by the QD fine-structure splitting (FSS)~\cite{Hudson2007a,Winik2017,Muller2018}:
	\begin{equation}
		\ket{\psi(t)}=\frac{1}{\sqrt{2}}\left(\ket{H_{XX}H_X}+\exp{\left(i\,\text{FSS}\,t/\hbar\right)}\ket{V_{XX}V_X}\right),
	\end{equation}
	where $\ket{H_{XX}H_X}$ ($\ket{V_{XX}V_X}$) describes a state where both photons are horizontally (vertically) polarized, $t$ the time between XX and X emission, and $\hbar$ the reduced Planck constant. With respect to the two-photon density matrix $\rho$ this temporal evolution manifests itself in a rotation (in the imaginary plane) of the outer non-classical entries $\rho_{VV,HH/HH,VV}$:
	\begin{equation}
		\rho = \frac{1}{2}\cdot\begin{pmatrix}
			1 & 0 & 0 & e^{-i\frac{\,\text{FSS}}{\hbar}t}\\
			0 & 0 & 0 & 0\\
			0 & 0 & 0 & 0\\
			e^{+i\frac{\,\text{FSS}}{\hbar}t} & 0 & 0 & 1
		\end{pmatrix}.
	\end{equation}
	
	\begin{figure*}
		\centering
		\includegraphics[width=1\linewidth]{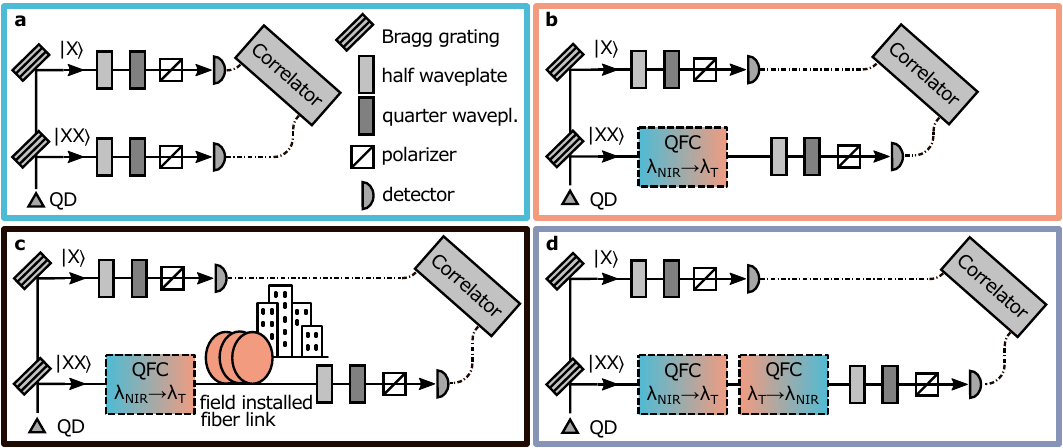}
		\caption{\textbf{Schematic drawing of the experimental configurations:} \textbf{a} The initial setup at \unit[780]{nm} before quantum-frequency conversion is shown. Volume-Bragg gratings filter the polarization-entangled exciton and biexciton emission lines into separate arms. A polarization-analyzing unit (quarter-, half-waveplate, polarizing beamsplitter and  fiber-coupled single-photon detector) in each arm together with a correlator allow to perform photonic tomography of the two-photon state.  \textbf{b} To be compatible with real-world telecommunication fiber networks a bi-directional polarization-preserving quantum frequency converter (QFC) is employed to convert the biexciton-to-exciton emission of the initial cascade from \unit[780]{nm} to \unit[1515]{nm}.  \textbf{c} The experiment is further expanded by sending this photon through a \unit[35.8]{km} field installed city fiber link. \textbf{d} A second bi-directional polarization-preserving frequency converter is utilized, bringing the \unit[1515]{nm} photon back to \unit[780]{nm}.}
		\label{fig:fig2}
	\end{figure*}
	\begin{figure*}
		\centering
		\includegraphics[width=1\linewidth]{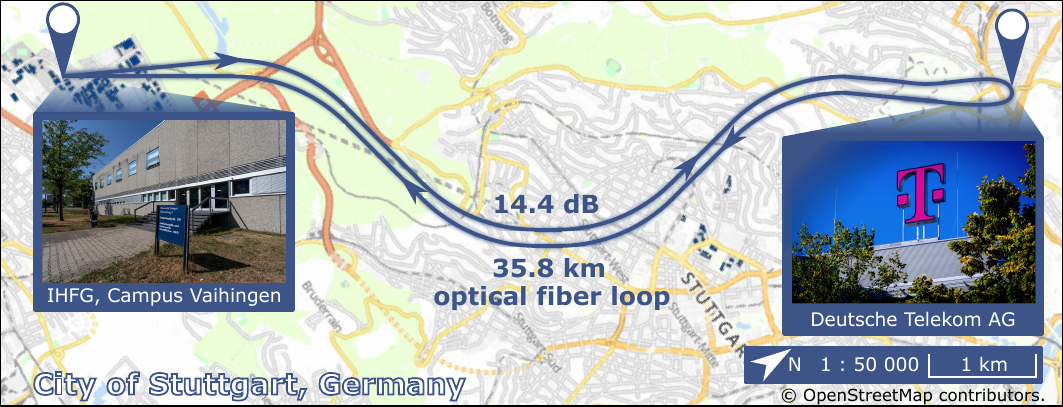}
		\caption{\textbf{Optical fiber course of the intra-city connection in Stuttgart, Germany, over which one photon of the entangled pair is transmitted:} In a research laboratory of the ``Institut für Halbleiteroptik und Funktionelle Grenzflächen'' (IHFG on the campus Vaihingen of the University of Stuttgart) an entangled-photon pair is created using a semiconductor quantum dot and the photon from the biexciton-to-exciton transition is frequency converted from \unit[780]{nm} to \unit[1515]{nm}. The frequency-converted photon is coupled into a standard single-mode telecom fiber, which is connected to a field installed city link. The photon is routed through the city of Stuttgart to a location of the Deutsche Telekom AG in Stuttgart Feuerbach. At this site it is then sent back to the research lab at the IHFG. The length of the entire optical fiber loop is \unit[35.8]{km} and has a \unit[14.4]{dB} fiber transmission loss. The shown static map image is taken from \textit{lageplan.uni-stuttgart.de} which is copyright OpenStreetMap, for further information see \textit{openstreetmap.org/copyright}.}
		\label{fig:fig3}
	\end{figure*}
\begin{figure}
		\centering
		\includegraphics[width=0.7\linewidth]{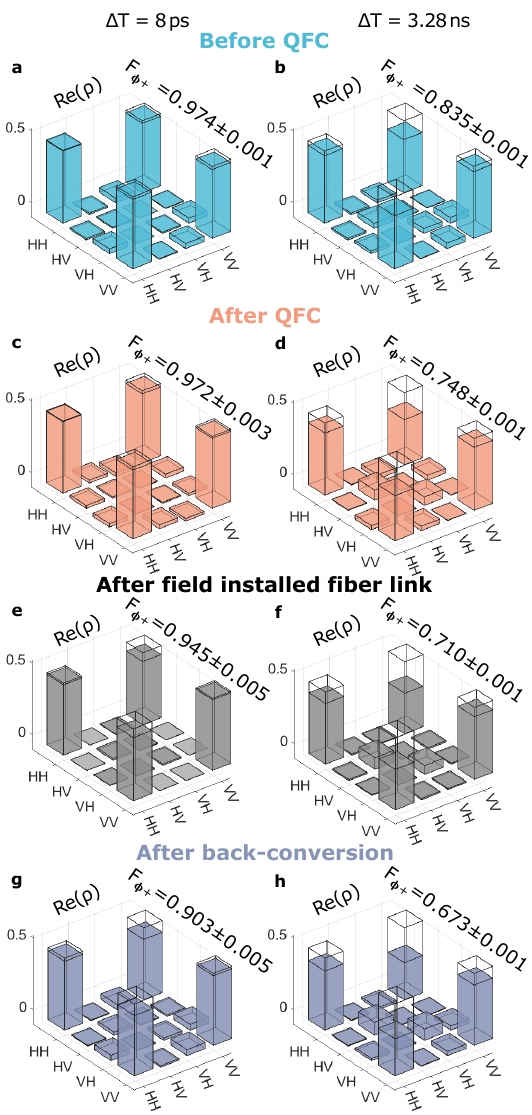}
		\caption{\textbf{Real part of the two-photon density matrix for different experimental configurations:} The measurements are evaluated for different integration time windows. In the left column data for an integration window of $\Delta T=\unit[8]{ps}$ (\textbf{a, c, e, g}) at zero time delay are displayed. The right column shows the density matrix over one laser repetition period ($\Delta T=T_{\text{rep}}=\unit[3.28]{ns}$, \textbf{b, d, f, h}). The state is depicted for photons before (\textbf{a, b}) and after (\textbf{c}, \textbf{d}) quantum frequency conversion (QFC). Furthermore, the experimental results for the photon transmission through an field installed city link (\textbf{e, f}) and the back-converted photon (\textbf{g, h}) are shown.}
		\label{fig:fig4ii}
	\end{figure}
\begin{figure*}
		\centering
		\includegraphics[width=1\linewidth]{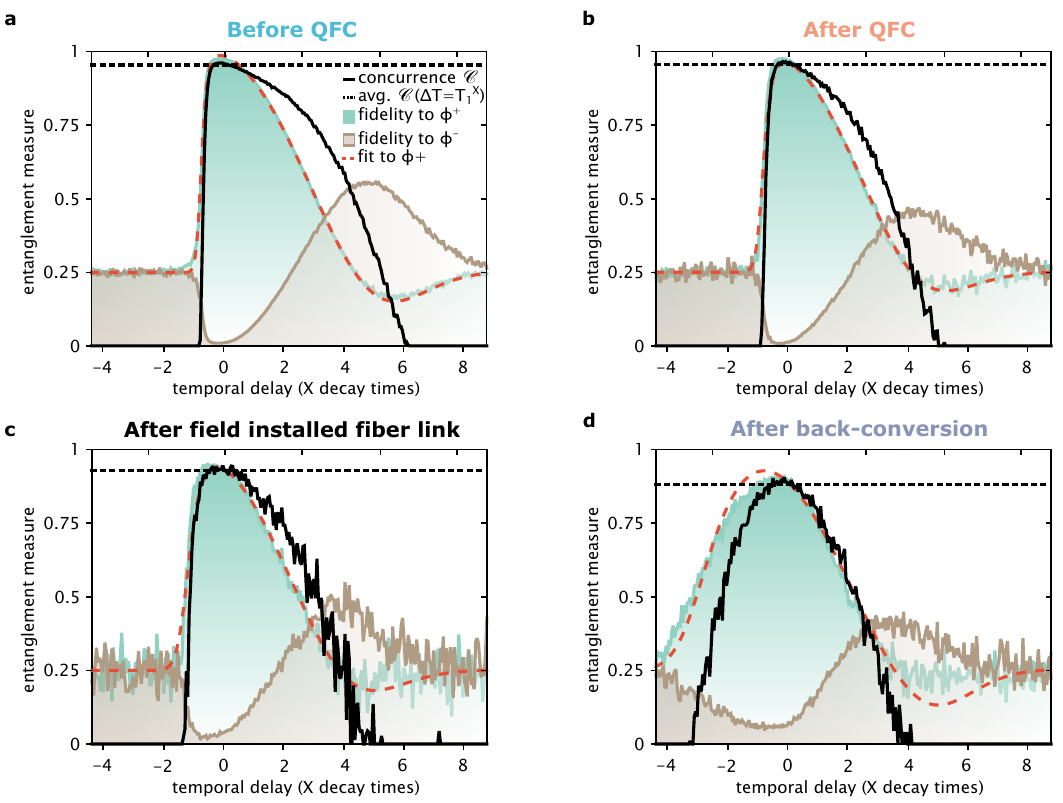}
		\caption{\textbf{Temporal evolution of entangled states under different experimental scenarios:} Various entanglement measures evaluated individually for different time delays in units of exciton decay times with a fixed temporal binwidth of \unit[8]{ps} are displayed. The fidelity to $\phi^+$ ($\phi^-$) is shown in green (brown). The fidelity to $\phi^+$ is fit with a model given in supplementary Eq.\,(S1) (dashed red line). As a fine-structure independent entanglement measure, the time dependent concurrence $\mathcal{C}$ (solid black line) and decay-time averaged concurrence $\mathcal{C}$ ($\Delta T = T_1^X$) (dashed black line) are depicted for the scenario before (\textbf{a}) and after (\textbf{b}) quantum frequency conversion, and after passing through the \unit[35.8]{km} field installed fiber link (\textbf{c}) and after back-conversion (\textbf{d}).} 
		\label{fig:fig5}
	\end{figure*}

	To verify the suitability of QD entangled photon sources for deployed experiments, the performance of the presented entangled-photon-pair source is investigated under four different scenarios with increasing experimental complexity (see Fig.\,\ref{fig:fig2}).\\
	First, to benchmark the degree of entanglement of the initial two-photon state, a photonic quantum-state tomography setup is implemented (Fig.\,\ref{fig:fig2}a)~\cite{Altepeter2005}. With an excitation rate of \unit[305]{MHz} a triggered fiber-coupled single-photon XX (X) count rate of \unit[507]{kHz} (\unit[450]{kHz}) is measured on the single-photon detector (detection efficiency $\eta_{\text{det}}=\unit[85]{\%}$). The differences in the two count rates stem from different fiber coupling efficiencies. More details about the resulting count rates can be found in supplementary Table\,S1-S4.\\
	Second, compatibility with present telecommunication networks is enabled by a high-efficiency polarization-conserving QFC unit (see supplements for details) converting XX photons from $\lambda_{\text{NIR}}=\unit[780]{nm}$ to telecom wavelengths at $\lambda_{\text{T}}=\unit[1515]{nm}$ ($\lambda_{\text{NIR}}\rightarrow\lambda_{\text{T}}$, Fig.\,\ref{fig:fig2}b) with a pump-laser wavelength of $\lambda_{\text{pump}}=\unit[1607]{nm}$. The final detector count rate for this part is \unit[44]{kHz}.\\
	Third, to probe the out-of-the-lab behavior of an entangled telecom photon, a transmission via an intra-city optical fiber loop is added (Fig.\,\ref{fig:fig2}c). In this scenario the employed fiber is a standard single-mode telecom fiber. As shown in Fig.\,\ref{fig:fig3} the photon is routed from the laboratory location at the ``Institut für Halbleiteroptik und Funktionelle Grenzflächen'' (IHFG on the campus Vaihingen of the University of Stuttgart) through the city of Stuttgart to a location of the Deutsche Telekom AG in Stuttgart Feuerbach. At this site it is send back to the research laboratory at the IHFG. The length of the entire optical fiber loop is \unit[35.8]{km} and has a \unit[14.4]{dB} fiber transmission loss, compatible with low absorption standard telecommunication networks. Supplementary Fig.\,S1 shows an optical time domain reflectometer (OTDR) measurement, indicating the looped and fusion sliced fiber link. The transmitted single photon count rate was \unit[1.35]{kHz}.\\ 
	Finally, to enable wavelength compatibility of an entangled telecom photon to quantum memories, like the well-studied $\text{Rb D}_{2}$ transition) or a spin-photon interface, a second QFC unit (instead of the field installed fiber link) is utilized to convert the telecom photon of the entangled pair back to NIR, to the Rb transition wavelength ($\lambda_{\text{T}}\rightarrow\lambda_{\text{NIR}}$, Fig.\,\ref{fig:fig2}d). In the conversion process from \unit[1515]{nm} back to \unit[780]{nm}, parasitical second harmonic light of the pump laser ($\lambda_{\text{pump}}=\unit[1607]{nm}$, $\lambda_{\text{pump,SHG}}=\unit[804]{nm}$) is generated in the QFC nonlinear crystal. Raman-scattering of the SHG light inside the crystal results in noise photons directly at the target wavelength of \unit[780]{nm} that can not be filtered spectrally~\cite{Kuo2018}. The SHG light itself is suppressed with a VBG ($\nu_{\text{VBG}}=\unit[200]{GHz}$) followed by a narrow spectral etalon filter ($\nu_{\text{etalon}}=\unit[0.9]{GHz}$). The filter window is below the inhomogeneous ($\nu_{\text{inhomo}}=\unit[5.6]{GHz}$) and homogeneous linewidth ($\nu_{\text{homo}}=\unit[1.3]{GHz}$) of the biexciton emission. While reducing noise photons, this strong filtering also alters the biexciton photon's temporal shape, by rendering it more symmetrical, reducing its signal strength (\unit[5.1]{kHz}).
	In all four configurations, photonic state tomography is performed to verify the preservation of entanglement.\\
	For this study, the density matrix $\rho$ of the two-photon state is extracted from polarization-resolved correlations employing a maximum likelihood estimation~\cite{James2001}. The polarization bases of the tomography unit need to be aligned with the polarization bases of investigated QD's dipole. A quick estimation of the entanglement allows to run an optimization routine, maximizing the estimated entanglement to find the correct polarization base alignment. While the first two experiments exhibit integration times of several hours, the latter two experiments are based on recorded data of several days. As no active polarization stabilization was employed, the two latter experiments were prone to minor polarization fluctuations (see supplementary Fig.\,S3 for an polarization stability measurement of the intra-city fiber link) on a timescale of half a day.\\
	Figure\,\ref{fig:fig4ii} shows the resulting  real part of the two-photon density matrix Re($\rho$) for all four experimental scenarios and two different integration windows of $\Delta T=\unit[8]{ps}$, and $\Delta T=\unit[3.28]{ns}$ which corresponds to the laser repetition period $T_{\text{rep}}$: In the left panels (Fig.\,\ref*{fig:fig4ii}a,c,e,g) $\rho$ is reconstructed from the integrated counts around an \unit[8]{ps} window of the correlation data at zero time delay. The \unit[8]{ps} time-window depicts the two-photon state right after state-initialization before FSS-induced state evolution starts to become relevant (FSS oscillation period $T_P=h/\text{FSS}=\unit[1.97]{ns}$). For the subfigure on the right (Fig.\,\ref*{fig:fig4ii}b,d,f,h), correlation data of an entire laser repetition period ($T_{\text{rep}}=\unit[3.28]{ns}$) are summed up to estimate $\rho$. This time window is relevant for applications utilizing the entire photonic wavepacket or when a resolution of the photon's temporal evolution is not feasible~\cite{Waks2002,Vyvlecka2023}. Long integration windows beyond an X decay time ($T_1^X=\unit[171]{ps}$) and FSS oscillation period ($T_P=h/\text{FSS}=\unit[1.97]{ns}$) will lead to the inclusion of additional noise photons and an averaging over the state-rotation thus reducing the entanglement.\\
	The following discussion will be of qualitative nature, while the exact measures of entanglement will be addressed in the examination of Fig.\,\ref{fig:fig5}. In Fig.\,\ref{fig:fig4ii}a it can been seen that the initial state right after preparation resembles a nearly pure $\phi^+$-state. Considering the entire photonic wavepacket ($\Delta T = \unit[3.28]{ns}$) Fig.\,\ref{fig:fig4ii}b shows a drop in the coherences $\rho_{VV,HH}$ and $\rho_{HH,VV}$, and a slight increase in the inner diagonals ($\rho_{HV,HV}$, $\rho_{VH,VH}$) together with a minor decrease in the two outer diagonals ($\rho_{VV,VV}$, $\rho_{HH,HH}$). These artifacts signify mixed components in the measured state (a fully mixed state corresponds to $\rho_{\text{mixed}}=\mathbb{1}/4$). Here, two effects contribute to this behavior: First, the large integration window includes uncorrelated noise photons. Second, the integration averages over the FSS-induced rotation resulting in an incoherent sum of maximally entangles states between $\phi^+$ and $\phi^-$ (weighted by the decay probability). This represents the baseline entanglement employed in the experiments. Examining the results in Fig.\,\ref{fig:fig4ii}c the conversion process does not change the appearance of Re($\rho$), validating preservation of entanglement after QFC. Our results after transmission through the intra-city fiber link and back-conversion of the photon shortly after preparation (Fig.\,\ref{fig:fig4ii}e,g) show only a slight decrease in the outer diagonals $\rho_{VV,VV}$, $\rho_{HH,HH}$ and increase in the inner diagonals $\rho_{HV,HV}$, $\rho_{VH,VH}$. We attribute these effects to multiple sources: A lowered signal-to-noise ratio (SNR) and polarization fluctuations in optical fibers add more mixed components (increasing inner diagonal elements) to the measured state. Furthermore, examining the corresponding complex values at zero time delay ($\rho_{VV,HH/HH,VV}^{loop}=0.44\pm i\,0.18,\,\rho_{VV,HH/HH,VV}^{back}=0.42\pm i\,0.18$, compare Supplementary Fig.\,S3) finite imaginary parts with opposing signs hint towards a state which is slightly rotated in the complex plane. These findings would support an unwanted rotation in the polarization-detection basis. We attribute the misalignment of the polarization-detection to a lowered SNR in the two latter experiments, making the alignment process relative to the QD basis (as explained above) more challenging. Overall, the demanding experimental condition of the intra-city fiber-network connection and back-conversion process have a mild effect on the measured entangled state proving the maturity of QD entanglement generation and QFC when applied in deployed environments.
	Studying Fig.\,\ref{fig:fig4ii}d,f,h, Re($\rho$) shares similar but more distinct features as Fig.\ref{fig:fig4ii}b with pronounced inner diagonals indicating mixed components. Here, transmission losses in the more complex experimental configurations and increased noise-levels from QFC units result in lowered SNRs: by expanding the integration window to $\Delta T=T_{\text{rep}}$ the measured signal contains a significant amount of noise photons, contributing to a more mixed state.\\
	To gain further understanding of the temporal dynamics of the emitted two-photon state, a more detailed analysis is carried out in the following. As a measure for entanglement, fidelity to two Bell states ($F_{\phi^+}$ and $F_{\phi^-}$) and the state-independent concurrence $\mathcal{C}$ are calculated for all four experimental conditions (Fig.\,\ref{fig:fig5}) from the estimated density matrices. As $\mathcal{C}$ is a measure for entanglement not bound to any specific quantum state it is unaffected by a FSS induced rotation, while the fidelity to a static state stays affected. The $F_{\phi^+}$ data are fit with a model given in supplementary Eq.\,(S1), taking into account the SNR - of the two-photon correlation as fixed parameter - and FSS. Here, state-of-the art fast single-photon detectors (Single Quantum EOS) enable a time-resolved analysis of the three measures and allow to follow the FSS-induced oscillation. The QDs fine-structure splitting of $\text{FSS}=\unit[2.1]{\mu eV}$ gives an oscillation period of $T_P=h/\text{FSS}=\unit[1.97]{ns}$ corresponding to $\unit{11.5}$ excitonic decay times ($T_1^X=\unit[171]{ps}$). For increasing temporal delays between X and XX emission the coincidence probability exponentially decreases lowering the SNR until the fraction of coincidences between X and XX photons approaches zero. Consequently, for longer temporal delays a damping in the measured state from a rotating pure two-photon state (oscillating between $\phi^+$ and $\phi^-$) towards a mixed state is expected. In Fig.\,\ref{fig:fig5} the entanglement measures are calculated for a given temporal delay in the two-photon correlation and are depicted with a temporal binning of \unit[8]{ps} (solid lines). Furthermore, the decay-time averaged concurrence (see supplementary Eq.\,(S3) and~\cite{Pennacchietti2024}) integrating over one excitonic decay time $\Delta T = T_1^X$ is shown (Fig.\,\ref{fig:fig5}, dashed black line).\\ 
	Starting with Fig.\,\ref{fig:fig5}a depicting the initial scenario, $F_{\phi^+}$ (green) increases rapidly from 0.25 (negative time delays) to a maximum value of $F_{\phi^+,init}^{max}=0.974\pm0.001$. For longer delays $F_{\phi^+}$ stays above 0.9 within the entire radiative X decay time, while it slowly descends below 0.25 reaching its minimum after $\sim$5 X decay times until it rises to a steady state at 0.25. $F_{\phi^-}$ (brown) shows an opposite behavior, dropping close to zero at zero time delay. For longer delays an increase to above 0.5 after $\sim$5 X decay times is followed by a drop back to 0.25. Comparing $F_{\phi^+}$ and $F_{\phi^-}$, they are $\sim180^{\circ}$ out of phase as expected. The oscillatory behavior in the fidelity is a signature of the described temporal evolution of the two-photon state. The fit to the data closely matches the described oscillation (see supplementary Eq.\,(S1)). The rotation from a $\phi^+$ into a $\phi^-$ state can be followed with a period beyond several X decay times such that the oscillation is dampened to a fully mixed state ($F=0.25$) after one period. As the switching rate between $F_{\phi^+}$ is slower than the X decay time, the maximum in $F_{\phi^-}$ can never reach the same level as $F_{\phi^+}$. The oscillation is slow enough that its impact on $F_{\phi^+}$ only becomes relevant after 0.73 X decay times, i.e. where $\mathcal{C}$ exceeds $F_{\phi^+}$. This slow oscillation results in a decay-time averaged fidelity of $0.9696\pm0.0007$. The concurrence $\mathcal{C}$ (black) shows a fast increase to $0.960\pm0.001$ for short time delays and drops slower (still above \unit{0.9} for \unit{1.3} X decay times) than $F_{\phi^+}$ reaching zero after \unit{6.2} X decay times. As $\mathcal{C}$ is unaffected by the FSS induced rotation it stays at higher values over an extended period of time eventually being lowered by the decreasing SNR as well. A decay-time averaged concurrence of $0.9544\pm0.0002$ is calculated. The instrumental response of the detection system ($\Delta t_{\text{IRF}}=\unit[58\pm10]{ps}$, average value of measured data recorded on several different days) is responsible for a finite slope in the rise (drop) of  $F_{\phi^+}$ ($F_{\phi^-}$) and $\mathcal{C}$ around zero temporal delay. For Fig.\,\ref{fig:fig5}b,c the qualitative behavior is similar to Fig.\,\ref{fig:fig5}a, while in Fig.\,\ref{fig:fig5}d the data are temporally more smeared out due to the strong spectral filtering employed when two QFC units are utilized. Furthermore the three scenarios with QFC units (b, c, d) have a higher noise level in $F_{\phi^+}$, $F_{\phi^-}$ and $\mathcal{C}$.\\
	Examining the resulting numbers, the maximum reachable fidelity $F_{\phi^+,init}^{max}=0.974\pm0.001$ together with the decay-time averaged fidelity (concurrence) reaching $0.9696\pm0.0007$ ($0.9544\pm0.0002$) of the employed QD is comparable with similar state-of-the-art entangled-photon-pair sources~\cite{DaSilva2021,Hopfmann2020,Pennacchietti2024}. This high degree of entanglement is conserved by the QFC process ($F_{\phi^+,conv}^{max}=0.972\pm0.003$) with a decay-time averaged fidelity (concurrence) of $0.9624\pm0.0007$ ($0.9546\pm0.0002$). Sending the XX photon through the intra-city fiber loop slightly decreases the measured entanglement fidelity to $F_{\phi^+,loop}^{max}=0.945\pm0.005$, the time averaged fidelity (concurrence) is $0.9305\pm0.0011$ ($0.9302\pm0.0033$). This decrease is mostly due to polarization fluctuations (see supplementary Fig.\,S3) and attenuation of the signal, both caused by the field-deployed optical fiber. Converting the photon back to \unit[780]{nm} leads to a modest decrease in entanglement ($F_{\phi^+,backconv}^{max}=0.903\pm0.005$) with a decay-time averaged fidelity (concurrence) of $0.8897\pm0.0011$ ($0.8824\pm0.0033$). The introduction of a second converter and strong spectral filtering lowers the photon flux, while Raman-scattered SHG light of the pump laser increases background noise. Decreased SNR in the two latter cases leads to slight misalignments of the tomography polarization basis, as the optimization routine introduced before requires entangled photon pairs as a signal. These misalignments can cause errors in the calculation of an entanglement fidelity to a certain state and even the concurrence, if the measurement basis are not orthogonal.\\
	Main contributor for the modest lowering of the entanglement fidelities appear to be polarization fluctuations in optical fibers on the timescale of days. Furthermore, misalignments in the polarization-analysis setup caused by decreased SNR especially in experiments including a field installed city fiber link or an additional back-conversion step can impact not only the fidelity but the concurrence as well. These challenges can be tackled by employing an active polarization stabilization, reducing losses through the field installed city link and a spectral filter in the back-conversion exactly matching the XX photon's inhomogeneous linewidth increasing the single-photon flux. Addressing the FSS-induced state rotation, piezoelectric strain tuning the QD sample lifting FSS~\cite{Lettner2021} or fast birefringent modulators compensating the temporal state evolution~\cite{Wentland2023} can overcome the limits in the fidelity to a certain state. However, some applications allow for finite FSS-induced rotations given they can be resolved temporally~\cite{Pennacchietti2024}.\\

	\section{Discussion}
	In conclusion, we demonstrate the preservation of triggered polarization entanglement shared between a photon-pair emitted by a QD device (emitting at \unit[780]{nm}) in several scenarios: i) the intrinsic quantum dot photon pair entanglement degree is benchmarked. ii) one photon of the entangled pair is frequency converted to telecom wavelengths (at \unit[1515]{nm}). iii) the frequency converted photon is transmitted via a \unit[35.8]{km} intra-city optical fiber loop. iv) the photon is converted back (instead of the fiber loop) to the near-infrared regime. The employed bi-directional polarization insensitive QFC process hardly changes the near-unity fidelity of $F_{\phi^+,init}^{max}=0.974\pm0.001$. Despite transmission losses comparable to real-world field installed fiber networks, we show high entanglement fidelities after triggered photon transmission from semiconductor QDs along a field installed \unit[35.8]{km} intra-city optical fiber link ($F_{\phi^+,loop}^{max}=0.945\pm0.005$). The second polarization-conserving QFC step maintains a significant degree of entanglement ($F_{\phi^+,back}^{max}=0.903\pm0.005$). These findings constitute a tunable quantum photonic interface for Rb quantum memories (D2 line) or GaAs QD spins. However, time-resolved measurements reveal concurrence values as large as i)~$0.9544\pm0.0002$, ii)~$0.9546\pm0.0002$, iii)~$0.9302\pm0.0033$, iv)~$0.8824\pm0.0033$ within an entire exciton decay for the four described scenarios respectively.\\
	Looking forward, our results solidify the way towards fiber-based, terrestrial, long distance entanglement distribution. Entanglement distribution is key for photonic quantum communication and cryptography~\cite{Schimpf2021,Zahidy2024}. Together with the results of Weber et al.~\cite{Weber2019c} our findings prepare the ground for solid-state based quantum-repeater architectures with entanglement swapping as one of the central technologies.
	
	\subsection*{Funding} This work is funded by the German Federal Ministry of Education and
	Research (BMBF) via the projects QR.X (Contracts No.~16KISQ013, 16KISQ001K and 16KISQ016) and Q.Link.X (Contract No.~16KIS0864). The field installed SASER TestNet fiber infrastructure is partly funded by BMBF 16BP17200, and EUREKA, CELTIC-PLUS CPP2011/2-5.
	
	\subsection*{Acknowledgments}
	The authors gratefully acknowledge the company Single Quantum for their persistent support. We also thank the Deutsche Telekom for providing the field installed SASER TestNet fiber infrastructure. We would like to warmly acknowledge the steadfast support of Dr. Sebastian Kiesel and Andreas Berg (University of Stuttgart Information Center) by enabling the fiber connection, on campus Vaihingen, from the laboratory up to the Telekom deployed fiber network.
	
	%
	
	\subsection*{Disclosures}
	\noindent The authors declare no competing interests.
	
	\subsection*{Data availability}
	\noindent Data underlying the results presented in this paper are not publicly available at this time but may be obtained from the authors upon reasonable request.

		\bibliography{deplEnt_16}

\end{document}